\input psfig.sty
\input epsf.sty

\def\mev{\,{\rm Me\kern-0.1em V}}
\def\gev{\,{\rm Ge\kern-0.1em V}}

\documentstyle[12pt]{article}

\begin{document}
\begin{center}
{\Large{\bf Improved Pseudofermion Approach for All-Point Propagators}}\\
\vspace*{.45in}
{\large{A. ~Duncan$^1$
and E. ~Eichten$^2$}} \\
\vspace*{.15in}
$^1$Dept. of Physics and Astronomy, Univ. of Pittsburgh, 
Pittsburgh, PA 15260\\
$^2$Theory Group, Fermilab, PO Box 500, Batavia, IL 60510\\
\end{center}
\vspace*{.3in}
\begin{abstract}
Quark propagators with arbitrary sources and sinks can be 
obtained more efficiently using a pseudofermion method with a
mode-shifted action. 
Mode-shifting solves the 
problem of critical slowing down (for light quarks) 
induced by low eigenmodes of the Dirac operator.
The method allows the full physical content of every gauge configuration
to be extracted, and should be especially helpful for unquenched QCD
calculations.
The method can be applied for all the conventional quark actions: 
Wilson, Sheikoleslami-Wohlert, Kogut-Susskind, 
as well as Ginsparg-Wilson compliant overlap actions.
The statistical properties of the method are examined and
examples of physical processes under study are presented.
\end{abstract}
\newpage

\section{Introduction}

  The development of more powerful computing platforms as well as improvements
in algorithms suggest that unquenched lattice QCD gauge configurations on
reasonably large lattices should become available in the not too distant future.
The generation of equilibrated and decorrelated dynamical gauge configurations
for lighter quark masses will nevertheless remain extremely costly in
computational terms, and elementary considerations of load-balancing suggest
that we should be willing to expend a correspondingly large computational
effort in extracting the maximum physical content from each available gauge
configuration. Hadronic correlators built from conventional quark propagators
(which give the quark propagation amplitude from a single source to all points
on the lattice, for example) evidently exploit only a fraction of the physical
information latent in each gauge configuration. In this paper we
explore the statistical properties and computational feasibility of an
alternative stochastic approach to generating quark propagators, one which
supplies simultaneously quark propagation amplitudes from any point on a
space-time lattice to any other (i.e. all-point propagators).   
The use of pseudofermion fields has been studied previously
\cite{fbpseudo,hlpseudo}; particularly for single light quark systems 
(e.g. heavy-light mesons). With these approaches, 
as the quark mass becomes light the convergence of pseudofermion Monte Carlo 
calculation suffers critical slowing down.
In this paper, we observe that for physical correlators (in momentum space)
the statistical noise problem resides mainly in its 
low momentum behaviour. 
We will find that the convergence can be greatly improved 
by separating off the lowest eigenmodes of the Dirac operator and 
treating them exactly while doing the Monte Carlo calculations 
with a mode-shifted action.

In Section 2 we review the essential 
properties of pseudofermion fields
which allow the computation of an all-point propagator. 
Although the detailed results presented in this paper primarily 
concern Wilson or 
clover-improved quark actions, we also indicate here how the method can
be applied to the computation of an all-point overlap operator. Of course,
the basic properties of pseudofermion fields will already be familiar 
to practitioners of the  hybrid MonteCarlo \cite{HMC}
or the Luescher multiboson method \cite{Luescher}. 
In Section 3 we describe in some detail the use of pseudofermion 
quark propagators in constructing a variety of hadronic correlators 
(corresponding to both two- and three-point hadronic Greens functions), 
such as disconnected parts contributing to form factors or to hairpin 
amplitudes for isoscalar mesons, etc. 
In Section 4 we describe
the statistical properties of hadronic correlators computed using pseudofermion
all point propagators. The availability of all-point propagators allows 
high-statistics evaluations of the full four-momentum structure of such
correlators.  Detailed autocorrelation studies reveal that the method
gives very accurate results (with rapidly decorrelating amplitudes with 
a time constant of a few heat-bath sweeps of the pseudofermion fields) 
for all but the lowest lattice momenta. 
The critical slowing down in these modes is related to the
presence of low eigenmodes of the hermitian Wilson-Dirac operator. 
Projection and shifting of a few low-lying modes turns out to be 
computationally straightforward, and finally in
Section 5 we show that such mode-shifted pseudofermion simulations allow 
accurate extraction of {\em all} momentum components of hadronic correlators.

\newpage
\section{Pseudofermion Actions and All-Point Quark Propagators}
 
 A pseudofermion action suitable for computing all-point quark propagators is constructed
from the quadratic form $Q$ defining the basic quark action of the theory:
\begin{equation}
\label{eq:qrkact}
  S_{\rm quark} = \sum_{a{\bf x},b{\bf y}}\bar{\psi}_{a{\bf x}}Q_{a{\bf x},b{\bf y}}\psi_{b{\bf y}}
\end{equation}
where $a,b$ are spin-color indices and ${\bf x,y}$ lattice sites. For the time-being, we shall consider
Wilson or clover-improved actions only (later, we return to the case of  Ginsparg-Wilson \cite{Gins}
 compliant overlap actions). Then the operator $H \equiv \gamma_5 Q$ is hermitian and we
 can form a positive-definite bosonic (``pseudofermion") action as follows
\begin{equation}
\label{eq:pfact}
  S_{\rm pf} = \sum_{a{\bf x},b{\bf y}}\phi_{a{\bf x}}^{*}H^{2}_{a{\bf x},b{\bf y}}\phi_{b{\bf y}}
\end{equation}
in terms of a complex bosonic pseudofermion field $\phi$. Averages of products of the pseudofermion
 field with respect to the Gaussian weight defined in Eq(\ref{eq:pfact}) yield inverses of  $H^2$ in the usual
 way:
\begin{eqnarray}
\label{eq:pfsim}
 <<F(\phi,\phi^{*})>>  &\equiv& Z^{-1}\int  d\phi d\phi^{*} F e^{-S_{\rm pf}(\phi,\phi^{*})}  \nonumber \\
    Z &\equiv&  \int  d\phi d\phi^{*}  e^{-S_{\rm pf}(\phi,\phi^{*})}  \nonumber \\
  <<\phi_{a{\bf x}}\phi_{b{\bf y}}^{*}>> &=& (H^{-2})_{a{\bf x},b{\bf y}}
\end{eqnarray}
Note that the Gaussian dependence of the pseudofermion action implies
\begin{equation}
\label{eq:avgact}
  <<S_{\rm pf}>> = N_{\rm col}N_{\rm spin}V = 12V
\end{equation}
where $V$ is the lattice volume (=number of lattice sites). This exact result is extremely useful in checking for
 errors in the simulation algorithm and in determining equilibration of an ensemble. Of course, the 
 quark propagator of interest in lattice QCD simulations is (up to a trivial $\gamma_5$ factor) $H^{-1}$,
 not $H^{-2}$. This can easily be achieved by a single additional ``dslash" operation applied to the
 conjugate pseudofermion field:
\begin{eqnarray}
\label{eq:qprop}
    \tilde{\phi}_{b{\bf y}} &\equiv& (\phi^{\dagger}H)_{b{\bf y}}  \nonumber \\
   <<\phi_{a{\bf x}}\tilde{\phi}_{b{\bf y}}>> &=& (H^{-1})_{a{\bf x},b{\bf y}} = (Q^{-1}\gamma_5)_{a{\bf x},b{\bf y}}
\end{eqnarray}
Note that separate pseudofermion fields are needed for each quark propagator as averages of
products of  (say) four {\em bosonic} pseudofermion fields  will produce contractions with the
wrong sign relative to the corresponding fermionic 4 quark amplitudes. 
   The simulation of averages of  the kind found in (\ref{eq:pfsim},\ref{eq:avgact},\ref{eq:qprop}) can be readily accomplished by heat-bath updates,
 due to the simple Gaussian dependence of the action on the fields. For either Wilson or clover-improved
actions, the dependence of the pseudofermion weight on the pseudofermion field at a specific lattice site
 ${\bf x}$ takes the form
\begin{equation}
\label{eq:locweight}
  e^{-S_{\rm pf}} \simeq e^{-\phi^{\dagger}_{a{\bf x}}A_{{\bf x}ab}\phi_{b{\bf x}}+4\kappa{\rm Re}(\phi^{*}_{a{\bf x}}v_{a{\bf x}})}
\end{equation}
 where $v_{a{\bf x}}$ is a complex spin-color vector assembled from the pseudofermion field at nearest and
 next-to-nearest sites, as well as appropriate gauge-link variables connecting to those sites. The 12x12 matrix
 $A$ is a multiple of the identity (specifically $A=1+16\kappa^2$) for Wilson actions, and a more complicated 
hermitian matrix assembled from clover gauge fields in the SW-improved case. In either case, the heat-bath 
update of $\phi_{\bf x}$ is readily managed by completing the square in (\ref{eq:locweight}). For the clover actions, the
 matrices $A_{{\bf x}ab}$ can be prediagonalized just once at the start of the simulation, and the resulting stored
 eigenvalues and eigenvectors used to quickly generate $\phi$ updates at each site according to the
 weight (\ref{eq:locweight}).

   The pseudofermion method can be readily generalized to study the all-point overlap operator arising from an overlap
 action satisfying the Ginsparg-Wilson condition \cite{Gins,ovlp}. Let $H$ be the hermitian Wilson-Dirac
 operator with suitably chosen overlap mass. The all-point overlap operator for arbitrary bare quark
 mass is trivially computable once all matrix elements of the  nonlocal operator $\epsilon(H)\equiv H\cdot(H^2)^{-\frac{1}{2}}$
 are obtained. Using an optimal rational approximation \cite{optimal}, this nonlocal operator can be
 written
\begin{equation}
\label{eq:optrat}
   \epsilon(H) \simeq (a_0+\sum_{m=1}^{N}a_{m}\frac{1}{H^2+b_{m}})H
\end{equation}
where $a_m$ are real and the $b_{m}$ are real positive. The number of pole terms $N$ needed for a given level
 of uniform accuracy over the spectrum of $H$ is related in a
 fairly straightforward way to the condition number of $H$, but typical studies of the overlap operator have
 used $10 <  N  < 100$. We shall return to this issue in Section 4, where we show that mode-shifting can
 be used to dramatically improve this condition number and reduce the number of poles needed. The
 needed all-point operator can clearly be obtained by an average of $N$ pseudofermion fields, $\phi^{(m)},m=1,N$: we begin
from the positive definite action
\begin{equation}
\label{eq:pfovlp}
  S_{{\rm pf,overlap}} = \sum_{m=1}^{N} \phi^{(m)\dagger}(H^{2}+b_{m})\phi^{(m)}
\end{equation}
and construct the desired combination of pole terms from a corresponding combination of  pseudofermion
 fields, averaged relative to the weight (\ref{eq:pfovlp}):
\begin{equation}
\label{eq:epsavg}
 \epsilon(H)_{a{\bf x},b{\bf y}} = <<\sum_{m}a_{m}\phi^{(m)}_{a{\bf x}}\tilde{\phi^{(m)}}_{b{\bf y}}>>+a_{0}H_{a{\bf x},b{\bf y}}\end{equation}
The computation of all-point overlap propagators 
will require a simulation within a simulation (analogously to the situation for single source overlap propagators, where
inversions within an outer conjugate gradient inversion are required, \cite{optimal}).

\section{Hadronic Correlators from All-Point Pseudofermion Quark Propagators}

 In general, the computation of multipoint hadronic correlators involving 
$n$ quark propagators can be reduced to convolutions of $n$ pseudofermion 
fields, rapidly computed by fast Fourier transform (FFT). 
In fact, using FFT we can easily construct a wide range of correlators 
involving both local and smeared operators. In this section we
shall illustrate this by considering a number of examples of physical interest.
We note here that it may be advisable to combine all-point with 
conventional fixed source (or sink) propagators (obtained, say, from a 
conjugate gradient inversion), as accurate results become increasingly 
difficult as the number of allpoint propagators used increases, due to the 
large condition number of $H$, as will become clear in subsequent sections. 
Furthermore, the projection methods described in Section 5 
become increasingly essential in reducing the statistical errors of 
the pseudofermion average as the number of independent all-point propagators 
increase.

\subsection{Local Two-point Correlators}

 The all-point propagators obtained by the pseudofermion technique can be used to construct
the complete momentum-space dependence of 2-point correlators of scalar or  pseudoscalar densities,
or  vector or axial-vector currents, while exploiting the full physical content of each gauge
configuration. For example, we may be interested in the full 4-momentum transform 
\begin{eqnarray*}
 \Delta_{PS-PS}(q)\equiv\sum_{x,y}e^{iq\cdot(x-y)}\Delta(x,y) 
\end{eqnarray*}
of  the 2-point
pseudoscalar correlator  $\Delta(x,y)$, given by
\begin{eqnarray}
 \Delta(x,y)&=& <0|T\{\bar{\Psi}(x)\gamma_{5}\Psi(x)\;\bar{\Psi}(y)\gamma_{5}\Psi(y)\}|0>  \\
            &=& -<\rm{tr}((Q^{-1}\gamma_{5})_{xy}(Q^{-1}\gamma_{5})_{yx})>  \\
            &=& -<<\sum_{ab}\phi_{xa}(\phi^{\dagger}H)_{yb}\chi_{yb}(\chi^{\dagger}H)_{xa}>> \nonumber  \\
            &=& -<<(\phi^{\dagger}H\chi)_{yy}(\chi^{\dagger}H\phi)_{xx}>>  \nonumber \\
            &=& -<<(\tilde{\phi}\cdot\chi)_{yy}(\tilde{\chi}\cdot\phi)_{xx}>>
\end{eqnarray}
where (10) represents a conventional operator VEV, the single  brackets in (11) refer to a functional 
average over gauge-fields,  and the double brackets in (12) imply averages  over gauge configurations
as well as a pseudofermion average for each gauge configuration to determine the all-point
propagators $Q^{-1}$ in (11). 
The full momentum-space correlator thus becomes an easily evaluated fast Fourier transform of products of pseudofermion fields:
\begin{eqnarray}
\label{eq:FFT2point}
\Delta_{PS-PS}(q)   = -<< \rm{FFT}(\tilde{\chi}\cdot\phi)(q)\rm{FFT}(\tilde{\phi}\cdot\chi)(-q)>>
\end{eqnarray}

 At this point, it may be of use to the reader to indicate the computational requirementsof these pseudofermion simulations (see \cite{dpy} for more details).
For a 6$^4$  lattice, a single heat-bath update of the two pseudofermion
fields $\phi,\chi$ requires 0.366 sec. on a 1.5 GHz Pentium-4 processor. The
convolutions and FFT operations required to obtain the desired four-momentum field
$\Delta_{PS-PS}(q)$ in (\ref{eq:FFT2point}) require an additional 0.024 sec. and are performed after every
2 heat-bath updates of $\phi,\chi$. Typically, a sufficiently accurate  pseudofermion average for $\Delta_{PS-PS}(q)$ 
was obtained from 20000 measurements, corresponding to 2.1 Pentium-4 hrs. 
For 10$^3$x20 lattices, a heat-bath update costs 3.30 sec, a measurement of $\Delta(q^2)$ requires
0.43 sec. (again, performed every 2 heat-bath steps), and final averages are taken from
7000 measurements,  corresponding to 6.8 Pentium-4 hrs. For comparison, the evaluation
of a conventional conjugate-gradient single-source propagator on a 10$^3$x20 lattice 
requires 1.2 Pentium-4 hrs.

\subsection{Smeared Meson Propagators}
 In lattice QCD, hadron spectroscopy is usually studied using smeared hadron sources/sinks to optimize the
ground-state signal in each hadronic channel. A typical multistate propagation amplitude  from Euclidean time
 $t_i$ to $t_f$ 
might therefore involve evaluation of the matrix
\begin{eqnarray}
\label{eq:multistate}
\lefteqn{{\cal M}_{\alpha\beta}(t_i,t_f) \equiv } \\
& &  \sum_{\vec{x}\vec{y}\vec{z}\vec{w}}f_{\alpha}(\vec{x})f_{\beta}(\vec{y})
 <0|T\{\bar{\Psi}(\vec{z}+\vec{x},t_i)\gamma_5 O_{\alpha}\Psi(\vec{z},t_i)
\bar{\Psi}(\vec{w}+\vec{y},t_f)\gamma_5 O_{\beta}\Psi(\vec{w},t_f) |0>\nonumber
\end{eqnarray} 
where $f_{\alpha}(\vec{x})$ are a set of spatial smearing wavefunctions, $\Psi(\vec{x},t)$ denotes the quark field at
 spatial (lattice) point $\vec{x}$ and time $t$, and $O_{\alpha}$ are the appropriate spin matrices for the desired
hadronic channels. 
The sums over $\vec{z},\vec{w}$ project the physical states onto zero-momentum.
Of course, delta function choices for $f_{\alpha}$ allow us to use local 
sources or sinks. In terms of pseudofermion  quark propagators
\begin{eqnarray}
\label{eq:multipf}
 {\cal M}_{\alpha\beta}(t_i,t_f)&\equiv& -\sum_{\vec{x}\vec{y}\vec{z}\vec{w}}f_{\alpha}(\vec{x})f_{\beta}(\vec{y})
 \{<<\tilde{\phi}_{\vec{w}+\vec{y},t_f}O_{\beta}\chi_{\vec{w},t_f}\tilde{\chi}_{\vec{z}+\vec{x},t_i}O_{\alpha}\phi_{\vec{z},t_i}>> \nonumber \\
 &-&N_f<<\tilde{\phi}_{\vec{z}+\vec{x},t_i}O_{\alpha}\phi_{\vec{z},t_i}\tilde{\chi}_{\vec{w}+\vec{y},t_f}O_{\beta}\chi_{\vec{w},t_f}>>\}
\end{eqnarray}
where the second line of (\ref{eq:multipf}) ($N_f$=number of flavors) is present only for isoscalar amplitudes where there 
is a nonvanishing loop-back (``hairpin") amplitude. The four-fold spatial sum in (\ref{eq:multipf}) can fortunately be reduced
to operations linear in the spatial volume $V_s$ of the lattice via the magic of Fourier transforms. Define  smeared
pseudofermion fields
\begin{eqnarray*}
  \hat{\phi}^{\alpha}_{\vec{x},t_i} &\equiv& \frac{1}{V_s}
\sum_{\vec{p}}e^{i\vec{p}\cdot\vec{x}}f_{\alpha}(\vec{-p})\phi_{\vec{p},t_i} \\
  \hat{\chi}^{\beta}_{\vec{y},t_f} &\equiv& \frac{1}{V_s}
\sum_{\vec{p}}e^{i\vec{p}\cdot\vec{y}}f_{\alpha}(\vec{-p})\chi_{\vec{p},t_f} \\
\end{eqnarray*}
where $f_{\alpha}(\vec{p})$ is the Fourier transform smearing function and $\phi_{\vec{p},t}$ represents the spatially 
 Fourier transformed (and time-sliced) pseudofermion field. Taking the isovector amplitude part of (\ref{eq:multipf}) for simplicity, one finds
 that this reduces to
\begin{equation}
\label{eq:isovec}
 {\cal M}^{\rm isovec}_{\alpha\beta}(t_i,t_f)= -<<(\sum_{\vec{x}}\tilde{\chi}_{\vec{x},t_i}O_{\alpha}\hat{\phi}^{\alpha}_{\vec{x},t_i})
\cdot(\sum_{\vec{y}}\tilde{\phi}_{\vec{y},t_f}O_{\beta}\hat{\chi}^{\beta}_{\vec{y},t_f})>>
\end{equation}
The isoscalar contribution, if present, is trivially obtained by interchanging the pseudofermion fields in an obvious
 way. 
\subsection{Three Point Functions - The Pion Form Factor} 
 The computation of more complicated correlators, such as the 3-point functions needed to
extract form factors, is greatly facilitated if we have all-point quark propagators at
our disposal. A typical example is shown in Fig.1, where a smeared meson created at time
$t_0$ propagates to time $t_1$, where spacelike momentum $q$ is injected at spacetime point
$(\vec{y},t_1)$ by an electromagnetic current, followed by propagation of the final-state
meson to time $t_2$, when it is removed by an appropriate smeared-sink operator.

\begin{figure}
\psfig{figure=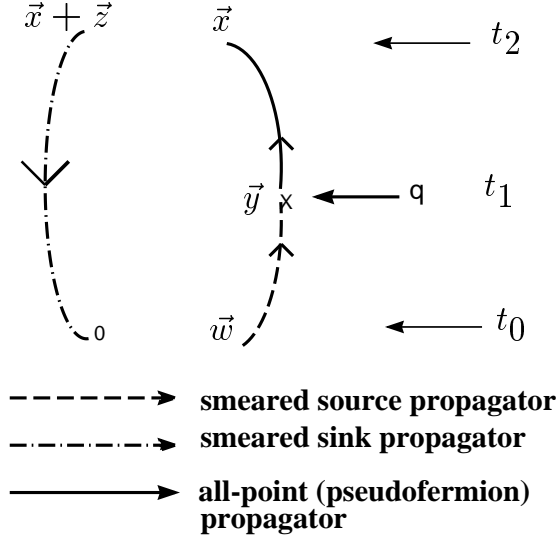,height=0.5\hsize}
\caption{Quark flow diagram for computing pion form factor(connected part)}
\end{figure}

The quark diagram displayed in Fig. 1 represents the connected contribution to the
following hadronic correlator:
\begin{eqnarray}
\label{eq:Jform}
\lefteqn{J_{t_{0}t_{1}t_{2}}(\vec{q}) = \sum_{\vec{w}\vec{x}\vec{y}\vec{z}}e^{i\vec{q}\cdot(\vec{x}-\vec{y})}f^{\rm sm}(\vec{z})f^{\rm sm}(\vec{w}) } \nonumber \\
&&\hspace{-0.5in} \cdot<\bar{\Psi}(\vec{z}+\vec{x},t_2)\gamma_5\Psi(\vec{x},t_2)\bar{\Psi}(\vec{y},t_1)\gamma_{0}\Psi(\vec{y},t_1)
\bar{\Psi}(\vec{w},t_0)\gamma_5\Psi(0,t_0)>
\end{eqnarray}
where in this case the same smearing function $f^{\rm sm}$  (optimized for a pion, say) is applied at both
initial and final times. The complete momentum-dependence of this 3-point function can be obtained by
using an all-point pseudofermion propagator for the quark propagator from $(\vec{y},t_1)$ to $(\vec{x},t_2)$.
The quark propagators into the point sink at $(0,t_0)$ and from the smeared source point at $(\vec{w},t_0)$
can be computed by conventional conjugate-gradient techniques. Again, by using Fourier transforms 
appropriately, the calculation of $J_{t_{0}t_{1}t_{2}}(\vec{q})$ can be reduced to operations at most linear
in the spatial lattice volume. Finally, as for the case of smeared meson propagators discussed above, the
disconnected contribution involving the contraction of $\Psi(\vec{y},t_1)$ back to $\bar{\Psi}(\vec{y},t_1)$
can be calculated with no extra effort, as the all-point propagator for just this case has been computed. In
 this case the pseudofermion field will appear in an average of the form $<<\tilde{\phi}_{\vec{y}t_1}\gamma_{0}\gamma_{5}\phi_{\vec{y}t_1}>>$.

\newpage
\section{Statistical Properties of Pseudofermion Propagators}
  In this section we shall describe the results of detailed studies of the statistical
properties of pseudofermion propagators computed from the basic formulas (3). We
first consider a case where only one all-point propagator is used in assembling the
full hadronic amplitude- namely, the 3-point correlator $J(q)$ giving the pion formfactor discussed
in Section 3.3. We shall see shortly that as a consequence of the typically high condition
number of the hermitian Wilson-Dirac operator $H$, autocorrelations for low-momentum
amplitudes become progressively longer as more all-point propagators are introduced
into the calculation (in Section 5, we shall show how to fix this problem by mode-shifting). Here,
we begin with an application where autocorrelations are relatively  unproblematic.

  For the pion form-factor calculation, we generated quenched configurations at $\beta=$5.9 on
12$^3$x24 lattices, and studied the resulting quark propagators at $\kappa=$0.1590 (with the
Wilson action). The results described here were obtained by studying the simulation of  the 
quantity $J(q)$ for a randomly chosen gauge configuration from this ensemble: examination
of other configurations reveals that the behavior we describe is generic.
From Eq(4) the (infinite ensemble) average pseudofermion action  
should therefore be $S_{\rm pf}=$497664. From Fig.2 it is apparent that the equilibrium 
value for this quantity is reached rather quickly, after about 20 heat-bath sweeps through the lattice.
An average of the value of 200 consecutive pseudofermion configurations after this gives
for example $S_{\rm pf}=$497599 with a standard deviation of $\simeq$650. Typically, we have performed measurements using
pseudofermion fields after 100 initial sweeps. 

\begin{figure}
\psfig{figure=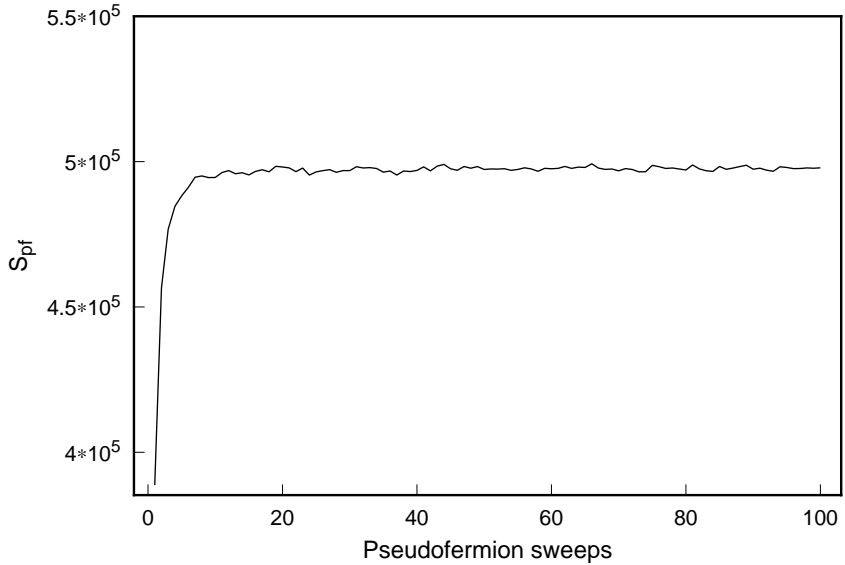,height=0.5\hsize}
\caption{Equilibration of pseudofermion average on 12$^3$x24 lattices} 
\end{figure}

  The decorrelation of hadronic amplitudes in the course of a MonteCarlo simulation of the
pseudofermion propagator (\ref{eq:pfsim}) is extremely sensitive to the particular momentum component
being calculated. In particular, low momentum components have a large overlap with the smallest eigenmodes
of $H$, which can have very small eigenvalues ($\simeq$10$^{-3}$ is not uncommon). These
low modes decorrelate only slowly in any local update of the action (2). This property
becomes immediately apparent when we examine the momentum dependence of either the convergence
of cumulative averages (Fig.3) of $J(q^2)$ (taking $t_2-t_1=t_1-t_0$=3) or the autocorrelation function
of the same quantity (Fig.4), as a function of number of pseudofermion heat-bath sweeps performed.
The autocorrelation time (defined as the integral under the autocorrelation curve of Fig. 4) turns out
to be about 13 for the zero-momentum mode and 7 for the $q^2=$1 (lattice units) mode: higher momenta
 (not shown here) yield autocorrelation times of order unity.  In fact, the autocorrelation times for this
correlator are fairly mild in comparison to the cases involving two all-point propagators discussed below:
8000 pseudofermion sweeps (= $\simeq$ 12 hrs on a 1.5GHz Pentium 4) allow  $J(q)$ to be extracted with
error bars well below the intrinsic fluctuation of the correlator from one gauge configuration to the next (see Fig.5).
\begin{figure}
\psfig{figure=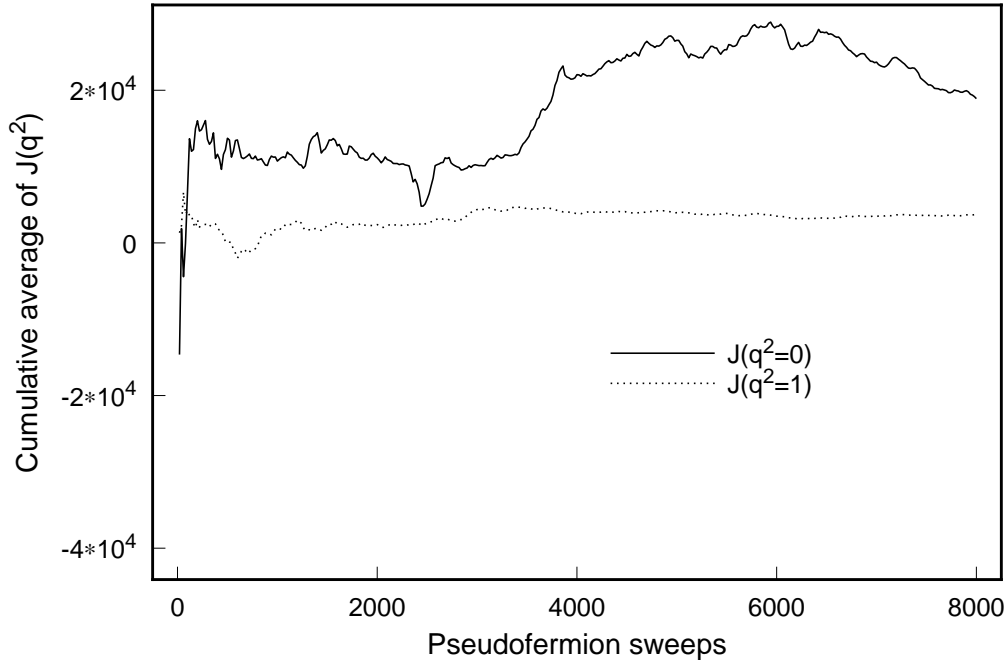,height=0.6\hsize}
\caption{Cumulative averages for pion form-factor amplitude $J(q^2),q^2=0,1$}
\end{figure}   

\begin{figure}
\psfig{figure=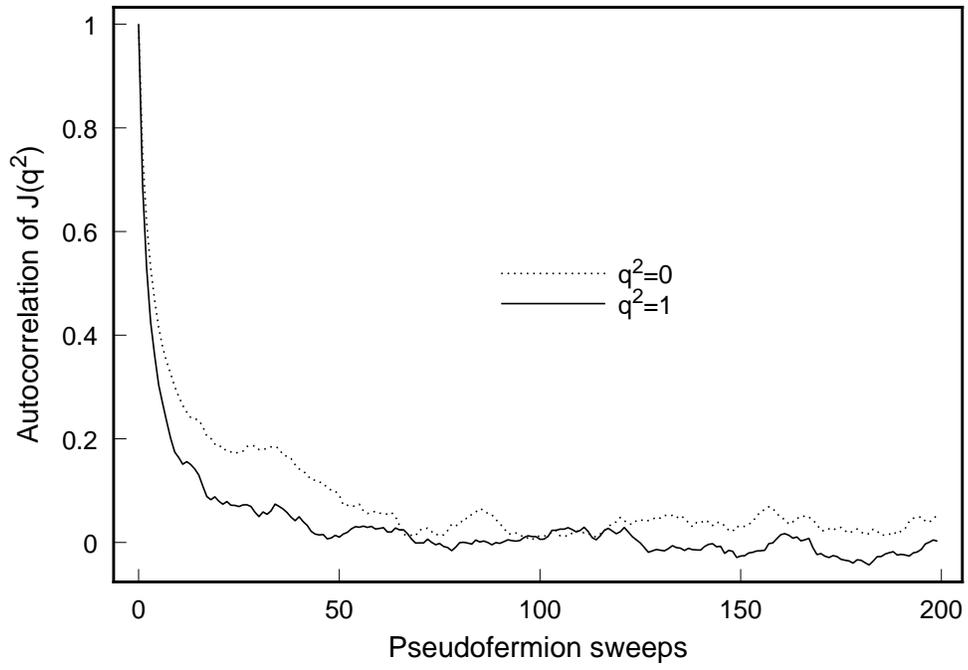,height=0.6\hsize}
\caption{Autocorrelation function of pion form-factor amplitude $J(q^2),q^2=0,1$}
\end{figure}   

  The critical slowing down seen in low eigenmodes becomes a more serious problem in situations
where two separate all-point propagators are required to construct the desired correlator, as in
the local two-point correlators discussed in Section 3.1. In this case the very large condition number
of $H^2$ implies an {\em extremely} slow convergence for the zero-momentum component of the
correlator. For example, from (11) we see that the local pseudoscalar density correlator can be
written, in momentum space
\begin{equation}
\label{eq:DPSPS}
  \Delta_{PS-PS}(q) = \sum_{x,y} e^{iq\cdot(x-y)} {\rm Tr}((H^{-1}_{xy}H^{-1}_{yx})
\end{equation}
which becomes, for zero momentum
\begin{equation}
\label{eq:DPS0}
  \Delta_{PS-PS}(0) = {\rm Tr}(\frac{1}{H^2})
\end{equation}
As one not uncommonly encounters small eigenvalues of $H$, it is apparent that a few low eigenmodes
can contribute disproportionately to this quantity. Moreover, these are exactly the modes that 
decorrelate most slowly in the pseudofermion simulation. To illustrate this, we have studied \cite{dpy} 
hadronic 2-point correlators on an ensemble of unquenched configurations generated with the
truncated determinant algorithm (TDA,\cite{tda}) on physically large coarse 6$^4$ lattices 
(lattice spacing $\simeq$0.4 fm). Cumulative averages for $\Delta_{PS-PS}(q)$ for a range of values of $q^2$,
for a typical gauge configuration in this ensemble, 
are shown in Fig. 5. The zero momentum mode is clearly not convergent even after 8000 pseudofermion
sweeps, while even the smallest nonzero (lattice) momentum component shows much more rapid convergence.
For the particular gauge configuration illustrated here, the lowest eigenvalue of $H$ turns out
to be 0.0024, which contributes 54\% of the total zero-momentum value ${\rm Tr}\frac{1}{H^2}$ !
The problem for this lowest mode can be seen (Fig. 6)
in another guise in the autocorrelation curves for $\Delta_{PS-PS}(q^2),0\leq q^2\leq 2$
(the autocorrelation time is $\simeq$ 1 pseudofermion sweep for $q^2>2$ (lattice units) so these curves are not
shown). Fortunately, the critical slowing down experienced in these pseudofermion simulations of
all-point propagators appears only to infect the very lowest momentum components. We shall see in the
next section that the problem can be eliminated for these components by mode-shifting a relatively small
number of low eigenmodes of $H$. This preconditioning substantially reduces autocorrelation times and
 allows us to extract reasonably accurate values even for the zero-momentum component of hadronic
correlators. The extraction and separate treatment of low eigenmodes is also essential in calculating
accurate all-point overlap operators.

\begin{figure}
\psfig{figure=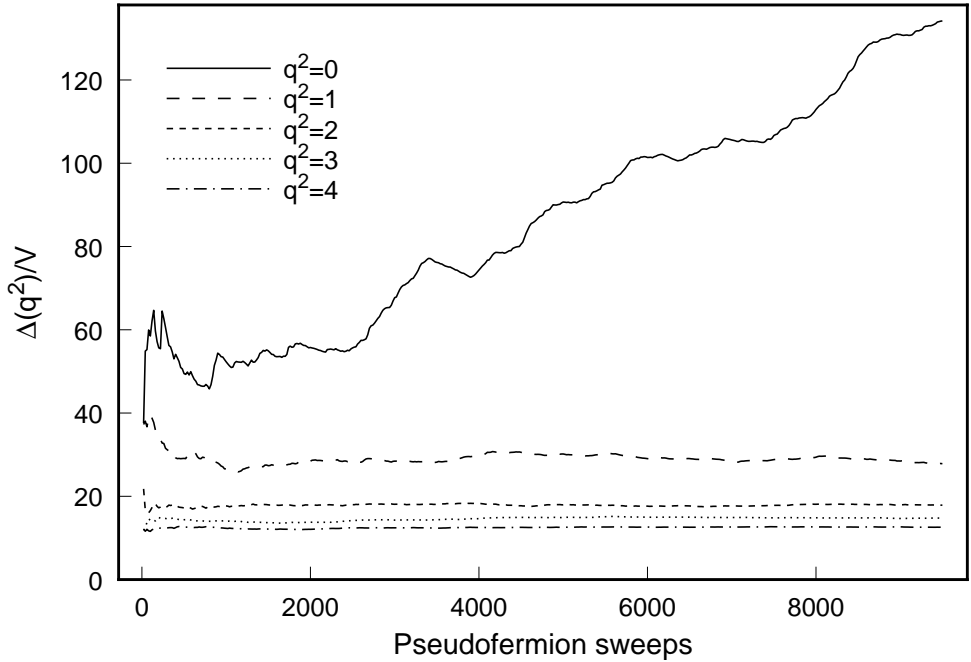,height=0.6\hsize}
\caption{Cumulative averages of pseudoscalar density correlator $\Delta_{PS-PS}(q^2)$, $0\leq q^2\leq 4$}
\end{figure}   

\begin{figure}
\psfig{figure=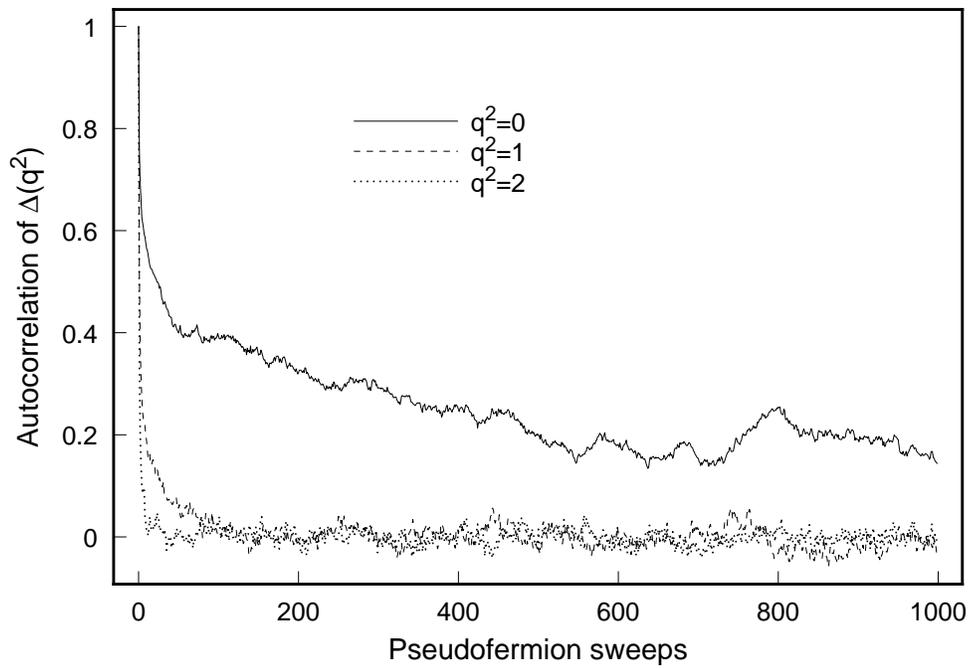,height=0.6\hsize}
\caption{Autocorrelation curves of  pseudoscalar density correlator $\Delta_{PS-PS}(q^2)$, $0\leq q^2\leq 2$}
\end{figure}   

\newpage
\section{Mode-Shifted Simulations of All-Point Propagators}

   The slow convergence of the lowest momentum modes discussed in the preceding section
can be substantially accelerated by shifting the low eigenmodes of the Hermitian 
Wilson-Dirac operator $H$ responsible for the critical slowing down. In the case of
the two-point correlator (\ref{eq:DPSPS}), the relevant parameter is the condition number of $H^2$,
which (for the ensemble of unquenched 6$^4$ lattices discussed in Section 4) can typically 
be reduced by two orders of magnitude by shifting the lowest 10 eigenmodes of $H^2$. More
generally, define
\begin{equation}
\label{eq:shifth}
  H_{s} \equiv H + \sum_{i=1}^{N} \delta_{i}{\bf v}_{i}{\bf v_{i}^{\dagger}}
\end{equation}
where ${\bf v}_{i}$ are a complete orthonormal set of eigenmodes of $H$, $H{\bf v}_{i}=\lambda_{i}{\bf v}_{i}$,
 and the lowest $N$ modes (in absolute value) are shifted:
\begin{equation}
\label{eq:deltai}
   \delta_{i} \equiv \lambda^{(s)}_{i}-\lambda_{i}
\end{equation}
For simplicity we shall take $\lambda^{(s)}_{i}={\rm sign}(\lambda^{(s)}_{i})
$ henceforth, although any value with magnitude  of order unity
will do. The extraction of low eigenmodes of $H$ is computationally straightforward using
implicitly restarted Arnoldi techniques \cite{ARPACK}: each mode requires a few minutes on
a Pentium-4 processor for the 6$^4$ lattices discussed here. 
  Corresponding to the shifted operator $H_{s}$ defined in (\ref{eq:shifth}) is a shifted pseudofermion action:
\begin{equation}
\label{eq:shiftpf}
  S_{\rm s,pf} = \sum_{a{\bf x},b{\bf y}}\phi_{a{\bf x}}^{*}(H_{s}^{2})_{a{\bf x},b{\bf y}}\phi_{b{\bf y}}
\end{equation}
Once the relevant low eigenmodes ${\bf v_{i}}$ are known, the heat-bath update of the pseudofermion
field $\phi$ can be performed by trivial modifications of the procedure outlined in Section 2. The
added computational load is not large: if $N$=10 modes are shifted, the time required for a pseudofermion
update increases by about 20\%. For the rest of this section, we shall use the double-bracket
notation  $<<....>>$ introduced in Section 2 to indicate averages relative to the weight generated
by the shifted action $S_{\rm s,pf}$. Accordingly, the {\em unshifted} quark propagator is given by
\begin{equation}
\label{eq:propform}
   H^{-1}_{a{\bf x},b{\bf y}} = <<\phi_{a{\bf x}}\tilde{\phi}_{b{\bf y}}>> - \sum_{i=1}^{N}\Delta_{i}{\bf v}_{i,a{\bf x}}\tilde{{\bf v}}_{i,b{\bf y}}
\end{equation}
where
\begin{equation}
\label{eq:Deltai}
  \Delta_{i} \equiv 1-\frac{1}{\lambda_{i}^{2}}
\end{equation}
and the tilde notation on the right-hand-side of (\ref{eq:propform}) still refers to the unshifted operator $H$, as in (\ref{eq:qprop}).
The pseudoscalar correlator in (10-12) can therefore be written
\begin{eqnarray}
\label{eq:finalDPSPS}
  \Delta_{PS-PS}(q) &=& \sum_{a{\bf x},b{\bf y}}e^{iq\cdot({\bf x}-{\bf y})}H^{-1}_{a{\bf x},b{\bf y}}H^{-1}_{b{\bf y},a{\bf x}}   \nonumber \\
   =\sum_{{\bf x},{\bf y}}e^{iq\cdot({\bf x}-{\bf y})}&\{&<<(\tilde{\phi}\cdot\chi)_{{\bf y}}(\tilde{\chi}\cdot\phi)_{{\bf x}}-\sum_{i}\Delta_{i}((\tilde{\chi}\cdot{\bf v}_{i})_{\bf x}(\tilde{\bf v}_{i}\cdot\chi)_{\bf y}+
(\tilde{\bf v}_{i}\cdot\phi)_{\bf x}(\tilde{\phi}\cdot{\bf v}_{i})_{\bf y})>> \nonumber \\
  &+&\sum_{i,j}\Delta_{i}\Delta_{j}(\tilde{\bf v}_{j}\cdot{\bf v}_{i})_{\bf x}(\tilde{\bf v}_{i}\cdot{\bf v}_{j})_{\bf y} \}
\end{eqnarray}
The term involving a double sum $\sum_{i,j}$ over shifted modes in (\ref{eq:finalDPSPS}) does not involve pseudofermion
fields and is therefore calculated just once. We see that the usual result (12) has to be supplemented by
an average of overlaps of the two pseudofermion fields with the shifted eigenmodes. Again, this is
computationally perfectly manageable.

  The dramatic effect of shifting even a few low eigenmodes on the convergence of correlators built from
pseudofermion averages is illustrated in Figs. 7,8,9,10.  As expected, the worst behavior is found in the
zero momentum mode, where the unshifted averages (Fig. 7) are still 60\% below the correct answer after 50,000
pseudofermion sweeps, while only 10,000 sweeps already give a reasonably good result after the 10 lowest modes
are shifted. The autocorrelation curves for the zero-momentum component tell the same story (Fig. 8): the
autocorrelation time is about 140 sweeps for the unshifted simulation, dropping to 11 (5) sweeps after shifting
10 (resp. 20) modes. The overall situation is much better for the lowest non-zero momentum mode, $q^2=1$,
as shown in Figs. 9,10. Here the autocorrelation times are roughly 8,3,2 sweeps for simulations with 0,10 and
20 shifted modes respectively, and the cumulative averages reflect a correspondingly higher stability.
The numerical evidence from these simulations clearly suggests that mode-shifting with $N$=10 modes, for the
ensemble of 6$^4$ TDA lattices considered here,  is
perfectly adequate for obtaining accurate results at all momenta.

\begin{figure}
\psfig{figure=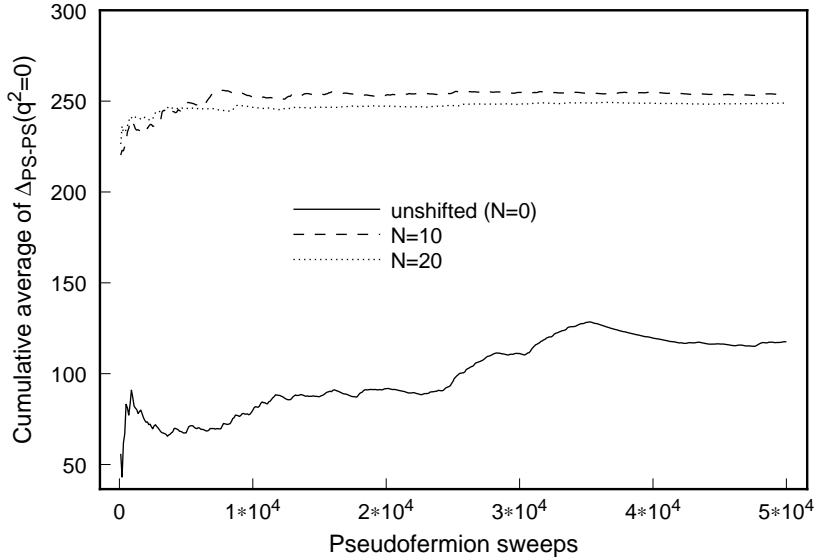,height=0.5\hsize}
\caption{Cumulative average of zero-momentum pseudoscalar correlator}
\end{figure}   

\begin{figure}
\psfig{figure=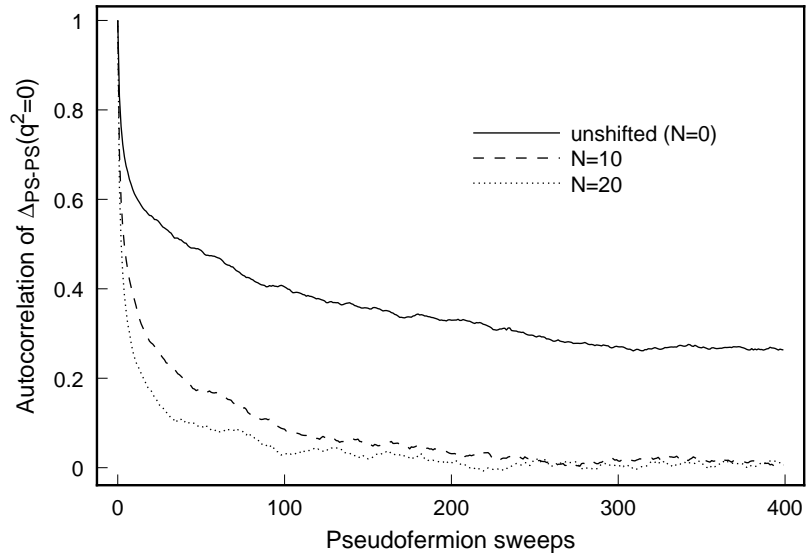,height=0.5\hsize}
\caption{Autocorrelation curve for zero-momentum pseudoscalar correlator}
\end{figure}   

\begin{figure}
\psfig{figure=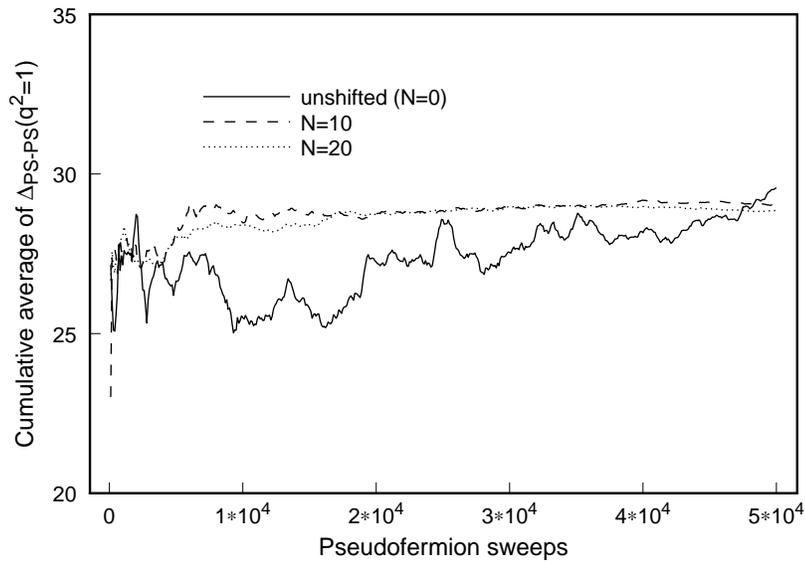,height=0.5\hsize}
\caption{Cumulative average for unit-momentum pseudoscalar correlator}
\end{figure}   

\begin{figure}
\psfig{figure=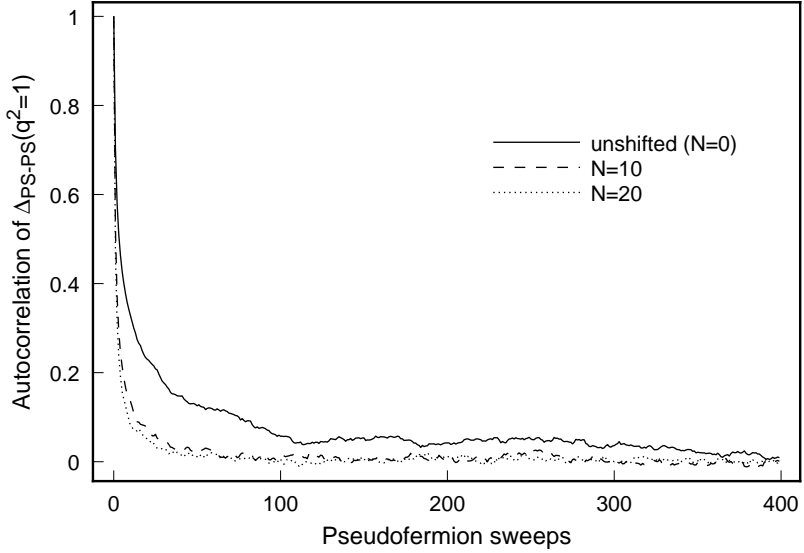,height=0.5\hsize}
\caption{Autocorrelation curve for unit-momentum pseudoscalar correlator}
\end{figure}   

  The calculation of an all-point overlap operator using (\ref{eq:optrat}) can be greatly facilitated by mode-shifting.
The number $N$ of poles needed in the optimal rational approximation to achieve a desired uniform accuracy
for $\epsilon(H)$ over the full spectrum of $H$ is directly related to the condition number (ratio of
highest to lowest eigenvalues) of $H$. On the other hand, provided the mode shifting preserves the algebraic
sign of the shifted eigenvalues (sign($\lambda^{(s)}_{i}$)=sign($\lambda_{i}$)), one clearly has 
$\epsilon(H_{s})=\epsilon(H)$. Accordingly, the condition number of $H$ can be drastically reduced by
computing the overlap pseudofermion operator using (\ref{eq:pfovlp}) with $H$ replaced with $H_{s}$ as the pseudofermion
action: the required code is identical to that used in the mode-shifted calculations of conventional
Wilson propagators described above.

\newpage
\section{Acknowledgements}
The work of A. Duncan  was supported in part by 
NSF grant PHY00-88946.
The work of E. Eichten was performed
at the Fermi National Accelerator Laboratory, which is
operated by University Research Association,
Inc., under contract DE-AC02-76CHO3000.

\end{document}